\documentclass{elsart}

\usepackage{graphics}
\usepackage{graphicx}
\usepackage{epsfig}

\usepackage{amsmath} 
\usepackage{amsfonts}
\usepackage{amssymb}

\def\lta{~\raise.4ex\hbox{$<$}\llap{\lower.6ex\hbox{$\sim$}}~}
\def\gta{~\raise.4ex\hbox{$>$}\llap{\lower.6ex\hbox{$\sim$}}~}
\def\ie{{\it i.e.}~}
\def\PD{\widetilde{\mbox{PD}}}

\begin{document}

\begin{frontmatter}


\title{On Evolutionary Spatial Heterogeneous Games}
\author[label1]{H. Fort}

\address[label1]{Instituto de F\'{\i}sica, Facultad de Ciencias, Universidad
de la Rep\'ublica, Igu\'a 4225, 11400 Montevideo, Uruguay}

\begin{abstract}
How coperation between self-interested individuals evolve is a crucial 
problem, both in biology and in social sciences, that is far from 
being well understood. 
Evolutionary game theory is a useful approach to this issue. 
The simplest model to take into account the spatial dimension 
in evolutionary games is in terms of cellular automata with just 
a one-parameter payoff matrix. 
Here, the effects of spatial heterogeneities of the environment and/or 
asymmetries in the interactions among the individuals 
are analysed through different extensions of this model. Instead
of using the same universal payoff matrix, bimatrix games in which 
each cell at site ($i$,$j$) has its own different `temptation to defect' 
parameter $T(i,j)$ are considered.
Firstly, the case in which these individual payoffs are constant 
in time is studied. 
Secondly, an {\it evolving} evolutionary spatial game such that 
$T$=$T(i,j;t)$, \ie besides depending on the position evolves 
(by natural selection), is used to explore the combination of 
spatial heterogeneity and natural selection of payoff matrices. 

\end{abstract}

\begin{keyword}
Complex adaptive systems \sep Evolutionary Game Theory \sep Cellular Automata

\end{keyword}
\end{frontmatter}


\section{Introduction}

Cooperation is ubiquitous in nature \cite{axel84} and essential 
for evolution \cite{ms97}.
How did cooperative behaviour evolve among self-interested individuals  
is an important open question in biology and social sciences. 
A powerful tool to analyse this issue is
evolutionary game theory \cite{ms82}-\cite{hs98}. It 
originated as an application of the mathematical theory of games to 
biological contexts, has become of increased interest to economists, 
sociologists, and anthropologists as well as philosophers.
Of particular interest are two-player games where each player has a 
strategy space containing two actions or `2$\times$2 games'.
In the problem of cooperation vs. competition the two actions or 
strategies are:
to cooperate (C) or to defect (D). The payoff of a player depends on 
its action and the one of its coplayer. Assuming symmetry between 
the two players, \ie that they are interchangable, there are four 
possible values for this payoff $R$, $S$, $T$ \&
$P$, corresponding respectively to the four situations 
[C,C], [C,D], [D,C] \& [D,D] (the first entrance correspond to the
action of the player and the second to the action of its opponent).
The paradigmatic example is the {\it Prisoner's Dilemma} (PD) in which 
the four payoff are ranked as $T$$>$$R$$>$$P$$>$$S$.
So $T$ is the `temptation' to defect, $R$ is the `reward' for 
mutual cooperation $P$ the `punishment' for mutual defection and 
$S$ is the `sucker's payoff'.
Clearly it pays more to defect ($T$$>$$R$ and $P$$>$$S$) 
but the dilemma is that if both play D they get $P$ that is worse 
than the reward $R$ they would have get if they had played C. 
The PD is connected with two other {\it social dilemma}
games \cite{poundstone}, \cite{l83}:
When the damage from mutual defection is increased
so that it finally exceeds the damage suffered by being exploited,
{\it i.e.} $T$$>$$R$$>$$S$$>$$P$,
the new game is called the {\it chicken} \cite{rap66}.
This game applies thus to situations such that mutual defection
is the worst possible outcome for both players as it happens in most
of animal contests.
On the other hand, when the reward surpasses the temptation
\ie $R$$>$$T$$>$$P$$>$$S$, 
the game becomes the
{\it Stag Hunt} (SH) \cite{s04}.
There are several animal behaviours that have been described as stag hunts.
For example, the coordination of slime molds. When individual amoebae of  
{\it Dictyostelium discoideum} are starving, they aggregate to form one   
large body. Here if they all act together they can successfully reproduce,
however the success depends on the cooperation of many individuals.

Classical evolutionary game theory 
constitutes a mean-field approximation which 
does not include the effect of spatial structures of populations. 
Axelrod \cite{axel84} suggested to place the individual ``players"
in a two-dimensional spatial array playing with its 
neighbours. These cellular automata (CA) were explored by Nowak and 
May \cite{nm93} who use a simple one-parameter payoff matrix
specifying a game that is in the frontier between the PD and chicken.
They found that the spatial structure allows cooperators to build
clusters in which the benefits of mutual cooperation can outweight
losses against defectors. This enables the maintenance of 
cooperation in contrast to the spatially unestructured game where 
defection is always favoured. 
Different variations of this Nowak-May (N-M) model were studied
taking into account distinct aspects, like the changes introduced by 
modifying the update rule \cite{st98}, 
the effect of allowing voluntary participation \cite{sh02}, 
the dependence on the graph topology \cite{ak01},\cite{svs05},
the effect of noise and connectivity structure \cite{vss06}, 
the consequences of the environmental stress by requiring a minimum 
threshold to survive \cite{aff06}, etc.

Here my goal is to extend the N-M model in several directions with the 
aim to study the following issues.

\begin{enumerate}

\item {\it The consequences of an heterogeneous environment}. 
This can be modelled through a payoff matrix varying from place 
to place.
For example, in a more hostile region the reward for mutual 
cooperation
can surpass the temptation to defect leading thus to switching
from the PD to the SH, etc.

\item{\it The effect of asymmetries in the interactions 
between the agents}. Indeed, asymmetries in the costs and 
benefits of cooperating have been
mentioned as an important ingredient in the evolution of 
cooperation \cite{cb02}.

\item{\it The emergence of payoff matrices from the very
natural selection process}.
The determination of the ranking of the payoff values
to explain the results of experiments or field observations
is a far from trivial matter. For instance, in the case of animals
there is controversy whether the PD or chicken is the appropriate 
game \cite{hp95,ns99}. Or many circumstances 
that have been described as PD might also be interpreted as a SH,
depending on how fitness is calculated \cite{s04}. 
Moreover, experimental studies indicate that the payoff matrix
is not a constant for very simple individuals like viruses \cite{tc03}.

\end{enumerate}

Hence, the basic idea is, instead of using the same universal payoff 
matrix, to use heterogeneous {\it bimatrix}
games \cite{hs98}, \ie each player has it own particular payoff matrix.
I begin studying a first variant of this model, 
in which the cell payoff matrix is regarded as something 
completely external to the individual, corresponding to environmental 
properties.
These properties change from place to place but not with time (at least
not in the temporal scale of the organisms) and thus the matrix diversity
is kept fixed during the evolution.

Then I focus on a second variant in which the cell payoff matrix 
is assumed to reflect `phenotypic' properties, internal to organisms,
thus it evolves together with its strategy.

\section{The Model}

The N-M model consists in a cellular automaton in which cells 
represent the simplest possible agents:  unconditional
players versus its neighbours. That is, at each generation or
time step $t$ of the game, there are those who always play C and those who
always play D.
Different types of neighbourhoods can be considered. For example
the {\it von Neumann neighbourhood} with $z$=4 neighbour cells 
(above, below, left and right cells). 
The score $U$ of a given player is the sum of the payoffs it collects 
against its $z$ neighbours. The payoff matrix depends on just 
one parameter $T$ while the other three have fixed values: 
$R$=1, $S$=$P$=0.
The dynamic is synchronous: all the agents update 
their states simultaneously at the end of each lattice sweep
\footnote{It is known that synchronous actualisation of the state 
of a lattice can induce artificial effects. I studied then what 
happened when performing an asynchronous update \ie each
site is updated after comparing with their neighbourhood. I found
qualitatively the same results.}.
Natural selection is mimicked through the simplest 'Imitate the Best'
(in the neighbourhood) update rule \cite{sf07} : 
Each individual, after playing against its neighbours,
adopts the strategy of the most successful neighbour
or $msn$
(the one that collected the highest utilities $U$ in the neighbourhood
at this round).

In this work two variants of the N-M model are studied:

\begin{itemize}

\item 
{\it Heterogeneous non evolving temptation.}

\vspace{2mm}

As a first step let's consider the version with the temptation 
parameter $T$ varying from cell to cell as a uniform random variable 
fixed in time: $T(i,j)$, where $i$ and $j$ are the coordinates of the 
center of the cell.
Since $P$=$S$, when $T(i,j)>$1 
the payoff matrix of the cell at ($i$,$j$) is in the frontier between 
the PD and the chicken, let's denote this game by $\PD_1$. 
On the other hand, when $T(i,j)<$1
the game played by the cell is in the frontier between
the PD and the SH and let's denote it by $\PD_2$.

\vspace{2mm}

\item
{\it Heterogeneous evolving temptation}.

\vspace{2mm}

Next I concentrate on the case in which the temptation
, besides varying from cell to cell, evolves with time 
\ie we have a variable $T(i,j;t)$ depending on the cell coordinates 
as well as the time step or generation $t$. 
It starts, at $t$=0, as a uniform random variable
and then evolves by natural selection: at each generation
$t$ all the members of a neighbourhood adopt the payoff matrix of
the $msn$.

\end{itemize}

Different intervals [$T_{min}$,$T_{max}$] for
$T(i,j;0)$ are explored. $T_{max}$ is taken in [1,2.5] in steps 
of 0.1, and for each of these values two values of $T_{min}$ are 
considered: 0 and 1.
Since $T_{max} \ge 1$, the possible games are determined by $T_{min}$:
when $T_{min}$=1 the payoff matrix of every cell corresponds
to $\PD_1$. On the other hand, 
the case $T_{min}$=0 is less restrictive. 
In some cells the game can be $\PD_1$ and in others $\PD_2$.

Square lattices of sizes ranging from $100 \times 100$ to
$500 \times 500$, with periodic boundary conditions are used.

The initial configuration for strategies is half of the cells,
chosen at random, playing C and the other half playing D. The 
temptation $T(i,j;0)$ varies from cell to cell as a uniform 
random variable in the interval [$T_{min}$,$T_{max}$].

For each generation $t$ the average fraction of cooperators
$\langle$$c$$\rangle$ and, in the second variant, the average temptation  
$\langle$$T$$\rangle$ are computed until the steady state is reached
(typically, this takes between 500 to 1000 generations).   
The symbols $\langle . \rangle$ denote averages that are both 
spatial, over all
the lattice cells, and over 500 runs each starting from a different initial    
configuration (to ensure independence of the initial conditions).

\section{Results}

\subsection{Heterogeneous non evolving temptation}  

Both the $z$=4 {\it von Neumann neighbourhood} 
and the $z$=8 {\it Moore neighbourhood} were studied producing
qualitatively the same results. Therefore, all the results
presented in this work are for $z$=4. 

\begin{figure}[htp]
   \begin{center}  
     \includegraphics[height=8.0cm]{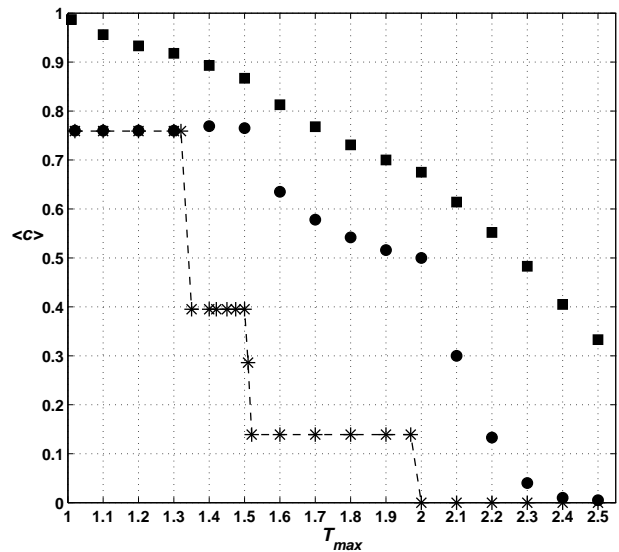}
   \end{center}
\vspace{-5mm}  
    \caption{Asymptotic value of $\langle$$c$$\rangle$ vs. $T_{max}$
for $z$=4: Nowak-May model (*), heterogeneous non evolving temptation 
model with
$T_{min}$=1 (filled circles) and $T_{min}$=0 (filled squares). 
Symbols are bigger than error bars.} 
\label{fig:cTmaxFix}
\end{figure}

Fig. \ref{fig:cTmaxFix} shows the asymptotic values of 
$\langle$$c$$\rangle$ vs. $T_{max}$. 
To allow comparison with the N-M model we also plot its results
(in this case the temptation is the same universal parameter 
$T$ for all the cells \ie  $T_{max}$ = $T_{min}$ = $T$). Roughly the 
$z$=4 N-M CA exhibits four different regions in the $T$ parameter 
axis, each corresponding to a given different asymptotic 
value of $\langle$$c$$\rangle$:   
1$< T \le$4/3, 4/3$< T \le$3/2, 3/2$< T <$2 and $T\ge$2.   
Note that the heterogeneous temptation model yields higher
values for the asymptotic fraction of cooperators.   
For $T_{min}$=1 it coincides with the N-M model
until $T_{max}$=4/3, reproducing its higher step, but then
it detaches from it.   
In the case of $T_{min}$=0 the differences are more drastic:
no steps are observed, and even for $T_{max}$=2.5 there's   
cooperation ($\langle$$c$$\rangle \simeq$ 1/3). 
In fact $\langle$$c$$\rangle$, as expected, is always greater 
for $T_{min}$=0 since, on average, the temptation to defect 
is smaller. For example, for  $T_{max}$=2, when $T_{min}$ 
changes from 1 to 0,  the fraction of cooperators 
jumps from  $\langle$$c$$\rangle \simeq$ 1/2 to
 $\langle$$c$$\rangle \simeq$ 2/3.

\subsection{Evolving heterogeneous temptation}

\begin{figure}[htp]
   \begin{center}  
     \includegraphics[height=8.0cm]{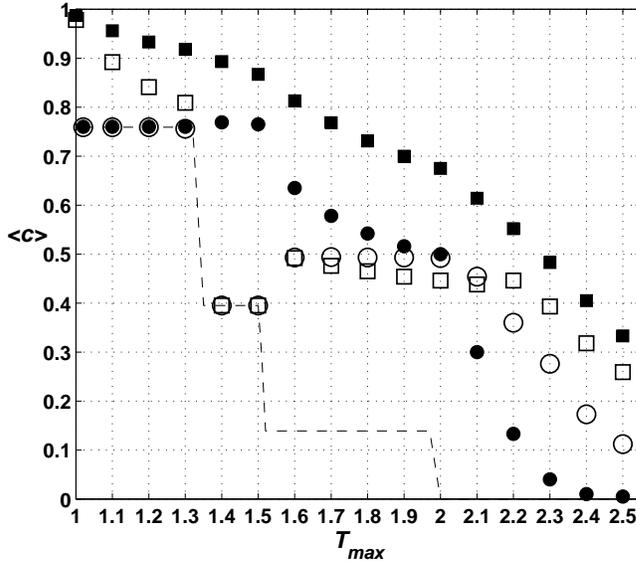}
   \end{center}
\vspace{-5mm}  
    \caption{The same as Fig. \ref{fig:cTmaxFix} plus
the asymptotic value of $\langle$$c$$\rangle$ vs. $T_{max}$
for the heterogeneous evolving temptation model with
$T_{min}$=1 (non-filled circles) \& $T_{min}$=0 (non-filled 
squares). Symbols are bigger than error bars.}   
\label{fig:cTmaxFull}
\end{figure}

Fig. \ref{fig:cTmaxFull} includes the data of Fig. \ref{fig:cTmaxFix} 
plus the corresponding asymptotic $\langle$$c$$\rangle$ produced
by the evolving temptation variant, for the same 
values of $T_{max}$ and $T_{min}$. 
Note that this version yields higher
values for the asymptotic fraction of cooperators.
For $T_{min}$=1 results are even more similar to the ones 
of the N-M model, there is complete coincidence 
up to $T_{max}$=3/2. When higher values of $T$ are allowed, 
something new occurs: the behaviour becomes non-monotonic,
there's also a step but now higher than the one to the left. 
This is quite unexpected, because when larger `temptation' 
values $\{T(i,j;0)\}$ are allowed, the fraction of cooperators 
becomes also larger. 
For $T_{max}>2$ the steps behaviour disappears.   
In the case of $T_{min}$=0 the points interpolate between the
ones of the corresponding heterogeneous non evolving variant 
and the N-M. In fact, they coincide with the points for
$T_{min}$=1 over the step [$T_{max}$=4/3, $T_{max}$=3/2]
and also show a non-monotonic variation for $T_{max}>$3/2. 

\begin{figure}[htp]
   \begin{center}  
     \includegraphics[height=10cm]{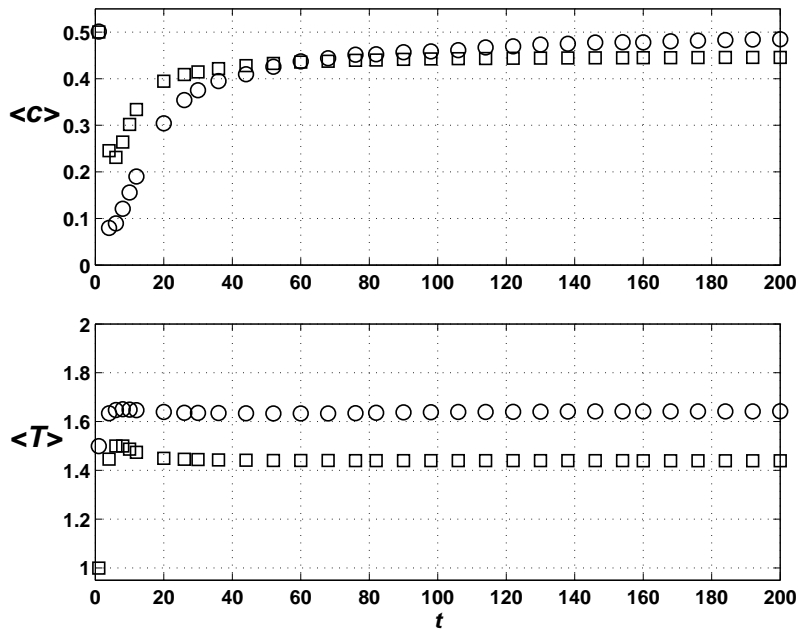}
   \end{center}
\vspace{-5mm}
    \caption{$\langle$$T$$\rangle$ (below) and 
$\langle$$c$$\rangle$ (above) vs. the generation $t$ (in both 
$T_{max}$=2) for:
$T_{min}$=1 (o) and $T_{min}$=0 ($\square$). 
Symbols are bigger than error bars.} 
\label{fig:cT}
\end{figure}

\begin{figure}[htp]
   \begin{center}  
     \includegraphics[height=10cm]{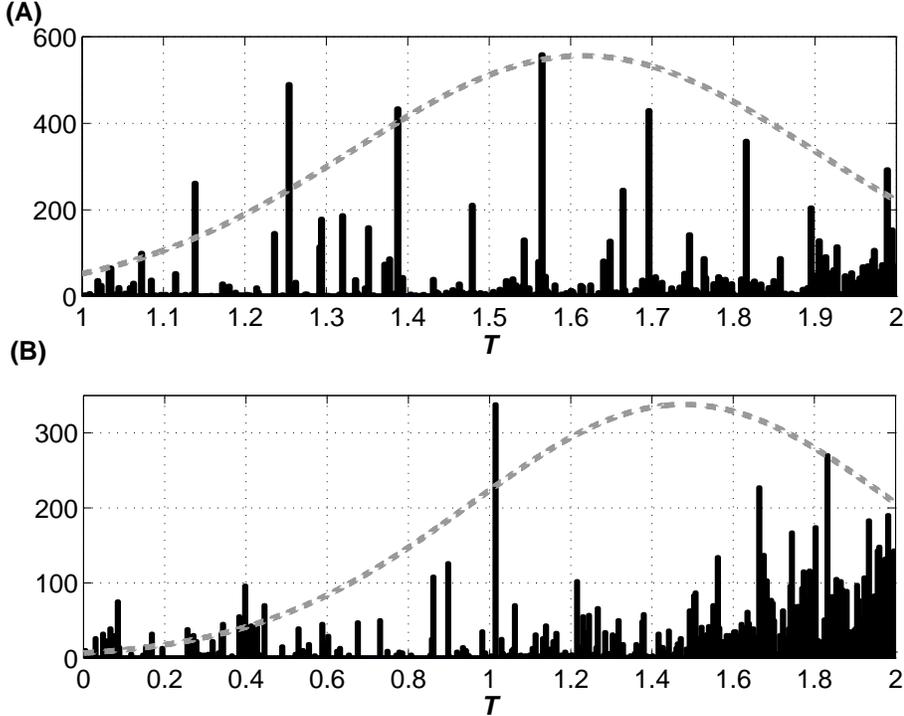}
   \end{center}
\vspace{-5mm}
    \caption{ Histograms for the asymptotic values of $T$ for 
   $T_{max}$=2: (A) $T_{min}$=1, (B) $T_{min}$=0. In gray: 
gaussian distributions for the same values of mean 
$\langle$$T$$\rangle$ and standard deviation $\sigma_T$.   }
\label{fig:histT}
\end{figure}

In Fig. \ref{fig:cT} the evolution of the average temptation 
$\langle$$T$$\rangle$ and fraction of cooperators 
$\langle$$c$$\rangle$ for $T_{max}$=1 and both for $T_{min}$=0 and 
=1 is shown. As expected, the value of 
$\langle$$T$$\rangle$ is lower for $T_{min}$=0 than for 
$T_{min}$=1.
Notice the sort of 'specular symmetry' between the curves of
$\langle T \rangle(t)$ and $\langle c \rangle(t)$ when they
evolve from 
$\langle$$T$$\rangle_0$=(2-0)/2=1 and $\langle$$c_0$$\rangle$=0.5 
, respectively, to their steady state values.
There is a short transient in which $\langle T \rangle$ 
($\langle c \rangle$) grows (drops) very quickly and then it 
decreases (increases) until it reaches its asymptotic value.   
These opposite behaviours are quite surprising since an increase 
of the average payoff for cheating is accompanied by an increase 
of the average cooperation level (just the opposite to what 
happens in the case
of non evolving temptation). 
Besides the average of the temptation $\langle$$T$$\rangle$, let's  
analyse also the evolution of its distribution. 
Fig.  \ref{fig:histT} shows the asymptotic (after 200 time steps)
distribution of values of the temptation $T$ for $T_{max}$=2 and
for $T_{min}$=1 (A) and $T_{min}$=0 (B), for a single initial 
condition. 
Hence, in both cases there is evolution from a uniform distribution 
to a non uniform one. The mean value for the temptation produced
by these histograms are, respectively, 1.615 and 1.42, both 
values in good agreement with the asymptotes for 
$\langle$$T$$\rangle$ in Fig. \ref{fig:cT}.  
In the case of $T_{min}$=0 a gaussian envelope can be completely 
discarded (gray dashed curve). For $T_{min}$=1 the situation is not 
so clear.

\begin{figure}[htp]
   \begin{center}  
     \includegraphics[height=18.0cm]{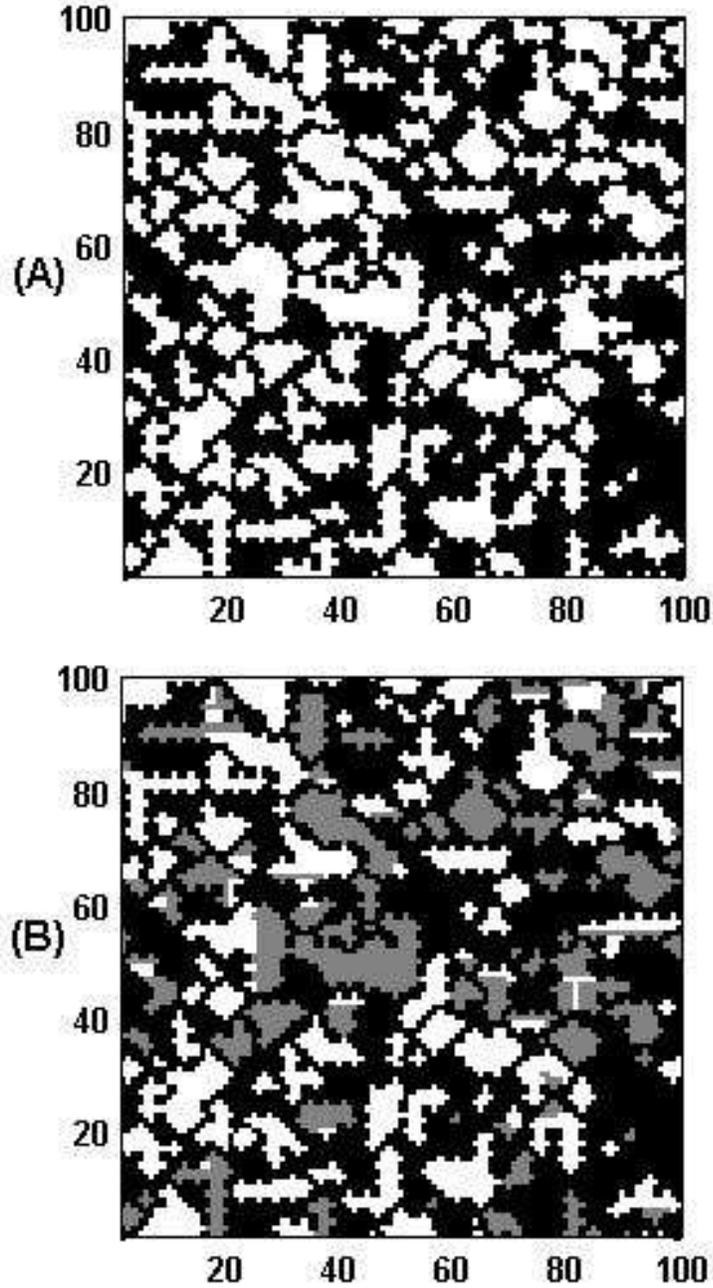}
   \end{center}
\vspace{-8mm}
    \caption{Steady state map of strategies $T_{max}$=2
             and $T_{min}$=0. 
             (A) C agents (white) and D agents (black). 
             (B) Agents with $T>1=R$: C (white), D (black); 
                agents with $T<1=R$ in grey.}
        \label{fig:Map}
\end{figure}

\begin{figure}[htp]
   \begin{center}  
     \includegraphics[height=10.0cm]{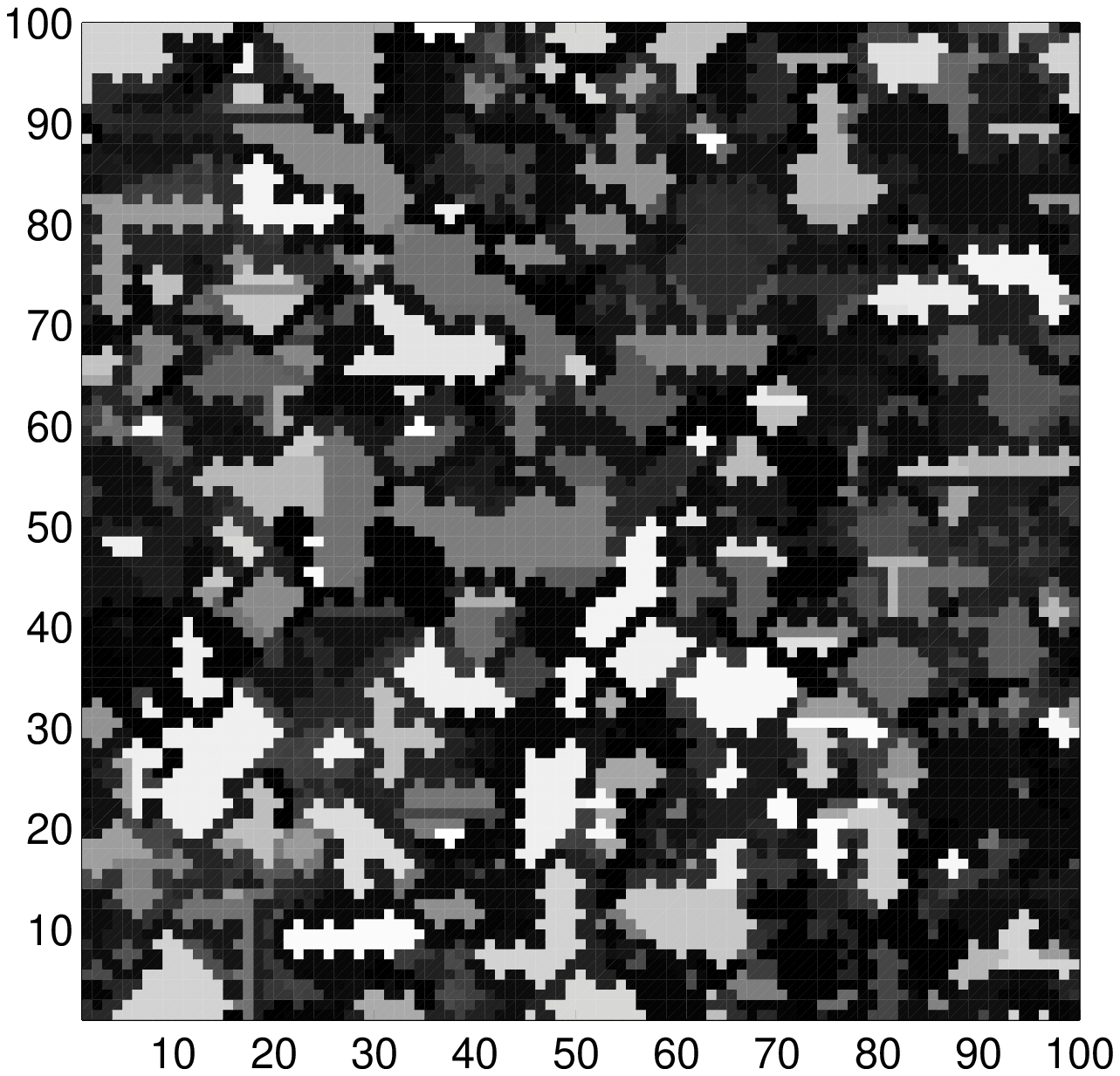}
   \end{center}
\vspace{-8mm}
    \caption{Steady state map of values of $T(i,j)$ for $T_{max}$=2
             and $T_{min}$=0. }
        \label{fig:MapT}
\end{figure}

The spatial patterns that emerge are illuminating.
For instance, let's analyse what happens for $T_{max}$=2
and $T_{min}$=0 for an arbitrary choice of initial 
conditions.
Fig. \ref{fig:Map}-(A) represents a 
typical steady state map showing 
clusters of C agents (white) on a 'sea' of D agents (black)
for a lattice of 100$\times$100=10,000 sites.
In Fig. \ref{fig:Map}-(B), all the agents that have a temptation 
$T<$1=$R$ are marked in grey. Notice that they are a subset of the C 
agents. In other words, all the agents that were 
selected with low values of $T$ 
are cooperators (the reciprocal is not true: many C agents have values
of $T>$ $R$=1). 
An explanation for this is that at the end only "successful" 
players remain \ie players whose strategies   
and phenotypes (temptation $T$) were copied by its neighbours. 
A player with a small $T$ clearly cannot be successful 
playing D so it must have been playing C.
 As a result of selection, the system evolves from an initial
configuration with $L\times L$ different payoff matrices 
(one per lattice cell)
to a situation in which many less matrices coexist. 
Starting in this case with 100$\times$100=10,000
payoff matrices one ends typically with around 500.  
This is shown in Fig. \ref{fig:MapT} which is the 
corresponding map of asymptotic values of $T(i,j)$. 
In this particular case it consists in 548 `patches' 
of different tones of gray, each corresponding to 
a selected value of $T$. 
One can recognize several of the domains that 
appear in Fig. \ref{fig:Map}-(B).
This is natural since the strategy (C or D) of the $msn$ 
is copied by agents together with its temptation. 
Also, it helps to understand why there are no clusters with white 
and gray sites well mixed in Fig. \ref{fig:Map}-(B): the typical 
length of the $T$-patches is much larger than the lattice spacing. 

The average individual score $\langle$$U$$\rangle$ can be approximated by
\begin{equation}
\tilde{U}  \simeq 0.5 \langle c \rangle ^2 +
\langle T \rangle \langle c \rangle (1- \langle c \rangle).
\label{eq:U}
\end{equation}
Substituting in equation (\ref{eq:U}) the computed values of
$\langle$$T$$\rangle$ and $\langle$$c$$\rangle$, one get a value 
that slightly overestimates $\langle$$U$$\rangle$.
For example, for $z$=4: 
$\tilde{U} \simeq$ 0.47 \&  $\langle$$U$$\rangle \simeq$ 0.43.

\section{Discussion and Final Comments}

Two variants implementing heterogeneous extensions of the N-M model
were introduced and studied. These model versions can be regarded as 
two extreme situations:
On the one hand, in the case of `temptations' $T(i,j)$ constant in time, 
we have the external environment point of view.
That is, the payoff matrix is regarded as something completely 
external to the individuals corresponding to the properties of 
physical background in which they are located.
These properties change from place to place but not with time 
(at least not in the temporal scale of the organisms) preserving thus
the original complete diversity of payoffs.
For example, in sites where the environment is more hostile the 
reward for mutual cooperation may become better than the 
temptation leading to switching
from the PD to the SH, etc. 
On the other hand, when the temptation evolves, we have the 
`phenotypic' point of view. In this case, one assumes that
the payoff matrix reflects properties of an organism, \ie is part 
of its phenotype, so it experiences natural selection.
Therefore the model can cope with diversity and asymmetric 
interactions between `players' either if they are product of an 
heterogeneous environment or of different `phenotypes' characterized 
by different temptations $T$. That is, for multiple causes, 
it is possible 
that the payoffs for you and your opponent are not equal (indeed this is  
what happens in general in real life).   
Furthermore, as it was mentioned, an empirical determination of 
the payoffs can be very difficult while variations in the payoff values can
dramatically alter theoretical predictions.
Here, one of my goals was to minimize the dependence of crucial model 
predictions, like the evolution of cooperation, on model parameters. 
So I have chosen the simplest agents -unconditional players 
(without memory or strategic parameters)- and I have not introduced 
a universal temptation parameter $T$ for all the players. 
Instead, starting with random heterogeneous   
distributions of $T$, I let that steady states with several 
domains characterized by different values of $T$ emerge
from the very process of natural selection.   
All the results are quite robust and do not depend on particular
payoffs choices, nor on the lattice topology or if the update   
is synchronous or asynchronous, and do not rely on 
specific characteristics of the agents. The dependence on the initial
conditions is also mostly removed by taking averages. 
Thus this really minimal model seems to combine robustness 
with realism and simplicity.

Both model variants yield a higher value of the asymptotic 
fraction of cooperators $\langle$$c$$\rangle$
than the one produced by the uniform N-M model.
It is worth remarking that for this heterogeneous payoff matrix model,
in general, $\langle$$c$$\rangle$ attains higher values for the the   
case of constant payoff matrices than for the evolving ones.
Also, for both model variants, in general, a diminution of $T_{min}$ 
(from 1 to 0) promotes a higher level of cooperation.
This seems natural since less competitive games (limit cases of SH)
are allowed.

The behaviour of $\langle$$c$$\rangle$ as a function of $T_{max}$=
is approximately monotonic for constant payoff matrices while in 
the case of evolving payoff matrices it exhibits non-monotonic 
variations. So, in this second variant, an interesting (unexpected) 
prediction is that by allowing larger initial `temptation' payoffs 
$\{T(i,j;0)\}$ the number of cooperators grows.

Concerning the emergence of spatial patterns,
the evolving temptation version organises into a
steady state that exhibits a rich structure with several 
'patches' of agents using the same payoff matrix. 
In addition, all the agents that reach the steady state 
with values of $T<R$=1 (when $T_{min}$=0) are cooperators.

In this work the evolution is driven just by natural selection.  
The effect of incorporating mutations, the other driving force of
evolution, is something important to explore. Work is in progress
on that direction. Nevertheless, preliminary results show that the
main conclusions presented here remain qualitatively the same.
It is also known that modifications of the update rule, 
and, in particular, an asynchronous dynamics may produce important changes
\cite{st98},\cite{svs05},\cite{sf07}. 
The effects of changing the update rule is something that deserves 
to be analysed separately in another work.

\vspace{2mm}

ACKNOWLEDGEMENTS

I would like to thank one of the referees who pointed out the work 
of M. Perc \cite{p06} on heterogeneous games and the effects of 
additive chaotic variations to the payoff matrix of the spatial 
PD game. The author proposes that this disturbances can arise
either by the player themselves or from the environment, situations
that correspond to the two different model variants I study here.

\end{document}